\documentclass[]{svjour3}

\usepackage{graphicx}
\usepackage{natbib}
\usepackage{amsmath}
\usepackage{amssymb}
\usepackage{color}

\newcommand{\ain}{{a_1}}
\newcommand{\aout}{{a_2}}

\newcommand{\zin}{{z_-}}
\newcommand{\zout}{{z_+}}

\newcommand{\elie}{{\mathbf E}}

\newcommand{\elik}{{\mathbf K}}

\newcommand{\asinh}{{\rm asinh}}
\newcommand{\mnras}{MNRAS}
\newcommand{\aap}{A\&A}
\newcommand{\apj}{ApJ}
\newcommand{\apjs}{ApJS}

\papertype{article}

\begin{document}

\title{Self-gravity in curved mesh elements}

\author{Jean-Marc Hur\'e \and Audrey Trova \and Franck Hersant}

\institute{Jean-Marc Hur\'e \and Audrey Trova \and Franck Hersant \at Univ. Bordeaux, LAB, UMR 5804, F-33270, Floirac, France\\ CNRS, LAB, UMR 5804, F-33270, Floirac, France\\ \email{jean-marc.hure@obs.u-bordeaux1.fr}\\\email{audrey.trova@obs.u-bordeaux1.fr}\\\email{franck.hersant@obs.u-bordeaux1.fr}}

\date{Received: date / Accepted: date}

\maketitle



\begin{abstract}
The local character of self-gravity along with the number of spatial dimensions
are critical issues when computing the potential and forces inside massive
systems like stars and disks. This appears from the discretisation scale where
each cell of the numerical grid is a self-interacting body in itself. There is
apparently no closed-form expression yet giving the potential of a
three-dimensional homogeneous cylindrical or spherical cell, in contrast with
the Cartesian case. By using Green's theorem, we show that the potential
integral for such polar-type $3$D sectors | initially, a volume integral with
singular kernel | can be converted into a regular line-integral running over
the lateral contour, thereby generalising a formula already known under axial
symmetry. It therefore is a step towards the obtention of another potential/density pair. The new kernel is a finite function of the cell's shape (with the
simplest form in cylindrical geometry), and mixes incomplete elliptic
integrals, inverse trigonometric and hyperbolic functions. The contour integral
is easy to compute; it is valid in the
whole physical space, exterior and interior to the sector itself and works in
fact for a wide variety of shapes of astrophysical interest (e.g. sectors of
tori or flared discs). This result is suited to easily providing reference solutions, and to reconstructing potential and forces in inhomogeneous systems by superposition. The contour integrals for the $3$ components of the acceleration vector are explicitely given.

\keywords{Gravity \and Disc \and Analytical methods \and Numerical methods \and Elliptic integrals}

\end{abstract}

\section{Main motivations}
\label{sec:itg}

As it is well known, Newton's inverse square law for gravitation between point
masses diverges at vanishing relative separations \citep{kellogg29,durand64}.
Consequently, in continuous media, the integral approach cannot provide
reliable potential values inside matter from simple quadrature schemes unless
{\it kernel singularities} are properly treated, either by direct integration
or through specific techniques. While the Poisson equation appears as the
privileged alternative, it requires the knowledge of accurate boundary conditions
that only the integral approach can provide, except in the very specific case of periodic boundary
conditions. The gradual increase in the precision of models and
simulations is a considerable source of motivation to derive new formulae, and
explore various techniques and algorithms to solve this problem more and more
efficiently \citep{gc01,maha03,hure05,jusu07,lietal08,gt11,mcj12}.

This article deals with the computation of the gravitational potential inside continuous systems, homogeneous or not, that can be discretised on a
polar-type $3$D-grid made of homogeneous cells/sectors. This includes cylindrical and spherical
grids which are commonly used in numerical models and simulations of bodies deformed by rotation and which exhibit a curved
shape, like stars and disks. Applications however exceed the astrophysical context. More
precisely, we show that the potential of an elementary sector of such grids is
always reducible to an one-dimensional, contour integral. \cite{DafaAlla20103624} have studied this issue from a finite difference approach. Our work extends a result already known under axial symmetry for closed toroids \citep[see e.g.][]{ansorg03}. Unfortunately, we failed to derive an algebraic, close-form expression for this contour integral. However, the
integral formula is relatively simple and general, and can serve to generating reference solutions or accurate potential values
at some points of space. From a purely practical point of view, the present formula is superior (accuracy and computing time) to the multipole approach which necessarily requires one triple integral over the material volume per term in the infinite series.

The paper is organised as follows. In Section \ref{sec:tb}, we recall Newton's
integral formula for curved, polar-type domains and set the notations and
hypothesis. In Section \ref{sec:lineinte}, we discuss the conversion of the
potential integral into a one-dimensional, contour integral through Green's
theorem. We briefly comment on the properties of the new kernel. We then apply
the result to the cylindrical cell; this is the aim of Section
\ref{sec:cylcell}. Explicit formulae as well as several examples are given. An appendix contains a few formulae and a demonstration. A basic Fortran 90 program is available as a supplementary resource linked to the online version of the paper (see Appendix \ref{app:f90}). Specific computations are also possible upon request. Expressions for the $3$ components of the acceleration are given in the Appendix \ref{sec:accs}.

\section{Theoretical background}
\label{sec:tb}

A volume ${\cal V}$ of space containing a mass density $\rho$ generates at a point P$(\vec{r})$ a scalar potential \citep{kellogg29}:
\begin{equation}
\psi(\vec{r})=-G\int_{\cal V}{\frac{\rho(\vec{r}') d^3 \tau}{|\vec{r}-\vec{r}'|}},
\label{eq:psi}
\end{equation}
where $\vec{r}'$ refers to points P$'$ belonging the source, $d^3 \tau$ is the elementary volume, $G$ for Gravitation problems. Successive integrations of the Green function $1/|\vec{r}-\vec{r}'|$ over the three spatial directions can sometimes lead to an algebraic formula \citep[e.g.][]{macmillan1930theory,durand64,binneytremaine87}, but this is rare. Then the Green function, which hyperbolically diverges inside sources, is usually expanded into spherical harmonics. In cylindrical coordinates where P$(R,\theta,Z)$, P$'(a,\theta',z)$ and $d^3 \tau=adad\theta' dz $, this expansion takes the classical form \cite[e.g.][]{ct99}:
\begin{equation}
\frac{1}{|\vec{r}-\vec{r}'|} = \sum_{m=-\infty}^\infty{e^{im(\theta-\theta')} \int_{t=0}^\infty{J_m(tR)J_m(ta) e^{\pm t z}dt}}
\label{eq:expansion}
\end{equation}
where $J_m$ is the Bessel function and $m$ is integer. In the absence of any symmetry, the potential is therefore given, at any point of space, by an infinite series where each term is a quadruple integral. From a computational point of view, this is obviously prohibitive. Besides, the truncation of the series also generates an error that adds to that due to the quadrature scheme. Many terms in the series | tens or hundreds| are necessary \citep{hachisu86,stonenorman92,mm12}. As often quoted in the literature, the oscillatory character of Bessel functions together with their range of definition are not well suited for numerical calculus. For all these reasons, we easily understand that such an expansion is not especially recommended, even if there is often no way around it. This remark is not specific to Eq.(\ref{eq:expansion}) and holds for spherical harmonics as well.

\begin{figure}
\centering
\includegraphics[width=14cm,bb=0 0 772 579,clip==]{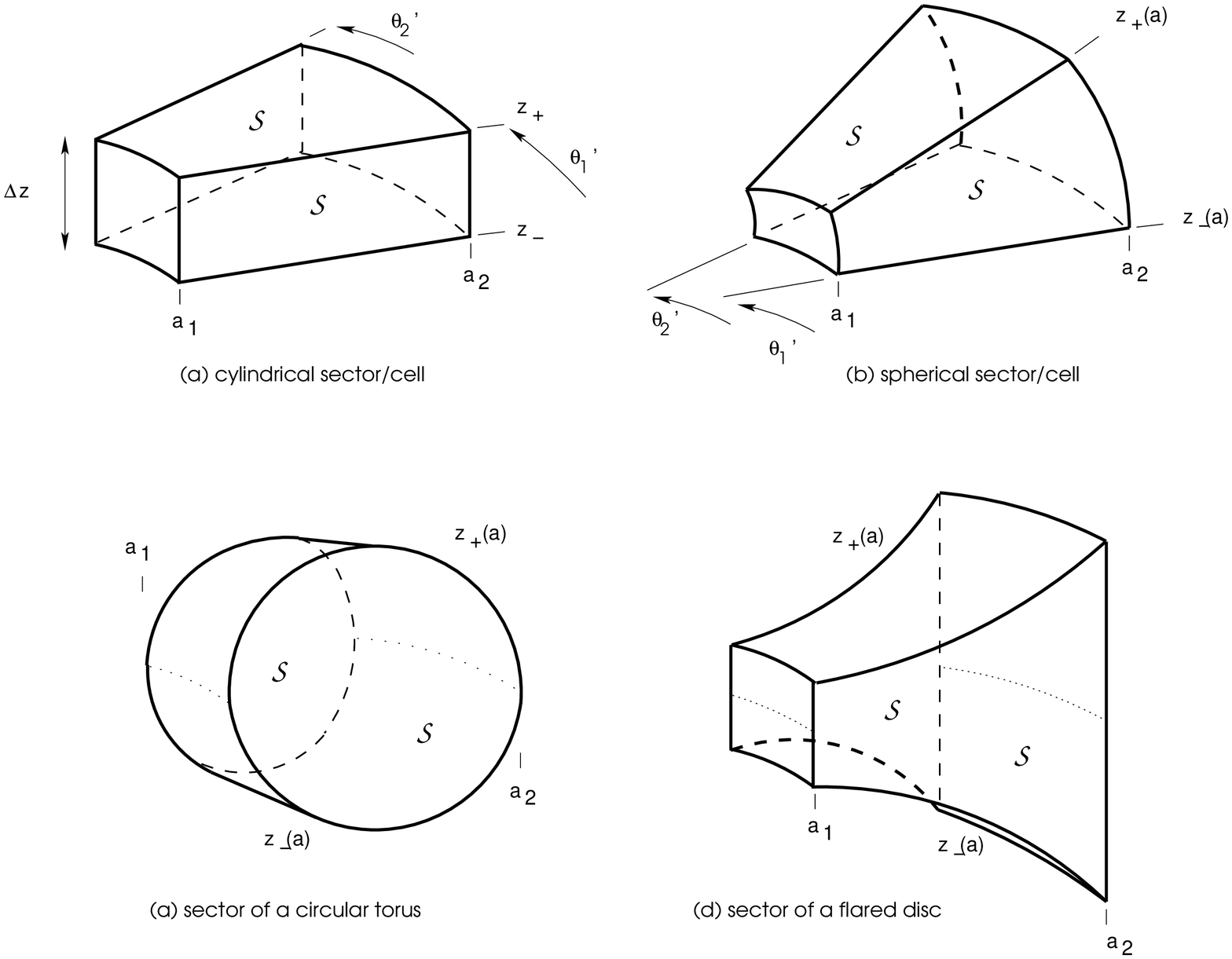}
\caption{Examples of possible shapes of astrophysical interest. All sectors considered here are bounded by two meridional planes defined by the polar angles $\theta=\theta_1'$ and $\theta=\theta_2'$. The lateral surface ${\cal S}$ is defined by an equation of the form $z_\pm(a)$ where $a$ is the radius.}
\label{fig:possiblecells}
\end{figure}

In the case of homogeneous systems as considered here, the number of integrals can advantageously be lowered. As a matter of fact, the integration over the polar angle $\theta'$ in Eq.(\ref{eq:psi}) is not trivial but feasible \citep{durand64,pz05}, which renders for instance Eq.(\ref{eq:expansion}) obsolete in the case where the system is $\theta'$-invariant (i.e., $\rho$ does not depend on $\theta'$ inside ${\cal V}$ whatever $a$ and $z$). Actually, we can rewrite Eq.(\ref{eq:psi}) as:
\begin{equation}
\psi(\vec{r})=-G\int_{\cal V}{\rho(\vec{r}')  \sqrt{\frac{a}{R}} \frac{k}{2} \frac{da d\theta' dz}{\sqrt{1-k^2 \cos^2\left(\frac{\theta'-\theta}{2}\right)}}},
\label{eq:psi2}
\end{equation}
where
\begin{equation}
\label{eq:kmod}
k = \frac{2\sqrt{aR}}{\sqrt{(a+R)^2+(z-Z)^2}}
\end{equation}
is a quantity in the range $[0,1]$.
Assuming that the volume ${\cal V}$ is a curve sector bounded by the two polar angles $\theta_1'$ and $\theta'_2$, then the above expression can be rewritten in terms of incomplete elliptic integrals of the first kind by setting:
\begin{equation}
\label{eq:beta}
\pi - (\theta'-\theta)=2\beta.
\end{equation}
So, we get:
\begin{equation}
\psi(\vec{r})=\int_{\cal S}{\rho(\vec{r}')\sqrt{\frac{a}{R}} k \left[ F(\beta_1,k)-F(\beta_2,k) \right] da dz},
\label{eq:psi3}
\end{equation}
where
\begin{equation}
F(\beta,k) = \int_0^\beta{\frac{dx}{\sqrt{1-k^2 \sin^2 x}}}
\end{equation}
is the incomplete elliptic integral of the first kind \citep{gradryz07,Olver2010}, $k$ is the modulus
defined by Eq.(\ref{eq:kmod}), ${\cal S}$ denotes the lateral face of the
material volume, $2\beta_1=\pi - (\theta_1'-\theta)$ and $2\beta_2=\pi -
(\theta_2'-\theta)$ according to Eq.(\ref{eq:beta}). We have now a bi-dimensional integral. Equation (\ref{eq:psi3})
applies to a wide variety of shapes, including cylindrical and spherical cells,
provided the mass density is uniform along any arc. An example of four possible
shapes of astrophysical interest is shown in Fig. \ref{fig:possiblecells}.
Note that the integrand in Eq.(\ref{eq:psi3}) is basically the potential of a
circular arc \citep{pz05}. The computation of $F(\beta,k)$ must be performed with
caution as soon as the amplitude $\beta$ stands outside the range
$[0,\frac{\pi}{2}]$ (see Appendix \ref{app:f}). In the axially symmetrical
case, we have for instance $\theta_1'-\theta=0$ and $\theta_2'-\theta=2\pi$,
and so, as expected \citep{durand64}, $F(\beta_1,k) -  F(\beta_2,k) =  2
\elik(k)$ where $\elik(k) \equiv F(\frac{\pi}{2},k)$ is the complete elliptic
integral of the first kind.

It is in principle possible to compute $\psi(\vec{r})$ from Eq.(\ref{eq:psi3})
for a given geometry by performing a double numerical integration over $a$ and
$z$. Not only this remains computationally costly, but there is also a
logarithmic singularity to manage (due to $F$) as soon as the amplitude
$\beta=\pi/2$ and $k=1$. If the body is a continuous pileup of polar cells
along the $z$-direction, then we can use the closed-form for the potential of
the polar cell \citep{hure12}. In this case, there is a single numerical integration to perform (see Sect. \ref{sec:accs}).

\section{Green's theorem}
\label{sec:lineinte}

It is clear from Eq.(\ref{eq:psi3}) that $dadz$ is an area element. There is therefore an alternative means since the surface ${\cal S}$ is, following our hypothesis, fully planar. Actually, the double integral can be converted into a line integral by using Green's theorem provided the kernel is the normal component of the curl of a vector $\vec{W}$, namely:
\begin{equation}
\int_{\cal S}  \vec{\nabla} \times \vec{W} \cdot d^2\vec{A} = \oint_{\partial \cal S} \vec{W} \cdot \vec{d\ell},
\end{equation}
where $d^2\vec{A} = d^2A \; \vec{n}$ is an area element oriented outward, and $d\vec{\ell}$ is an infinitesimal displacement counter-clock\-wise along the boundary $\partial \cal S$ of $\cal S$ in the plane $\perp \vec{n}$. In the present situation, the challenge is to find the appropriate vector $\vec{W}$ with cylindrical components $(M,0,N)$ such that:
\begin{equation}
\partial_a N - \partial_z M = \rho(a,z) \sqrt{\frac{a}{R}} k  \left[ F(\beta_1,k)-F(\beta_2,k) \right].
\label{eq:mandns}
\end{equation}
Clearly, $M$ and $N$ depend on $a$ and $z$. In this case, we have 
\begin{equation}
\int_{\cal S}{ \rho(a,z) \sqrt{\frac{a}{R}} k \left[ F(\beta_1,k)-F(\beta_2,k) \right] da dz} = \rho_0 \oint_{\partial \cal S}{Mda+Ndz},
\label{eq:curlt}
\end{equation}
which is therefore a one-dimensional integral (here, $\rho_0$ is a mass
density, introduced for homogeneity reasons).
 The existence of the two functions $M$ and $N$ is not guaranteed at all, not only due to the presence of
the special function $F$, but also because $\rho$ varies in general inside the
cell (except along any arc). In numerical simulations, the elementary cell is
precisely the smallest resolved pattern which marks the limit of the numerical
discretisation; in this context, the cell is, most of the time, considered as
a homogeneous system, i.e. $\rho=\rho_0$. This is the assumption we will make
in the following: $\rho$ is therefore constant in the whole integration domain
${\cal V}$.

As the potential depends on the difference of two incomplete elliptic integrals with particular amplitudes $\beta_1$ and $\beta_2$, we see that we can attack the problem by solving the generic problem:
\begin{equation}
\partial_a N_0 - \partial_z M_0 = \sqrt{\frac{a}{R}} k F(\beta_0,k),
\label{eq:mandns0}
\end{equation}
where $\beta_0$ is some amplitude and the two functions $M_0$ and $N_0$ are to be determined. In these conditions, the two functions $M$ and $N$ defined in Eq.(\ref{eq:curlt}) are given by:
\begin{equation}
\begin{cases}
M = M_1 - M_2,\\\\
N = N_1 - N_2,
\end{cases}
\end{equation}
where the subscripts $1$ and $2$ refer to the two amplitudes $\beta_1$ and
$\beta_2$, respectively. We could start the resolution of this problem by
considering a general expression for $M_0$ and $N_0$ on the basis of a linear
combination of the incomplete elliptic integrals, namely $F(\beta_0,k)$, the
incomplete elliptic integral of the second kind $E(\beta_0,k)$, and possibly
the incomplete elliptic integral of the third kind $\Pi(\beta_0,m^2,k)$ with
characteristic $m$. However, we are helped in this task by remembering that the
problem has already been solved under axial symmetry \citep{ansorg03}. Actually,
for a maximum opening angle $\Delta \theta' = \theta_2'-\theta_1'=2\pi$,
Eq.(\ref{eq:curlt}) is satisfied with the following components:
\begin{equation}
\begin{cases}
M=\zeta\sqrt{\frac{a}{R}} k \elik(k),\\\\
\label{eq:mnaxi}
N=\sqrt{\frac{a}{R}} k \left\{ (a+R)  \elik(k) - \frac{2R}{k^2} \left[  \elik(k) -   \elie(k) \right] \right\},
\end{cases}
\end{equation}
where $\zeta=Z-z$ and $\elie(k)=E(\frac{\pi}{2},k)$. We see the great advantage of this formulation: both $M$ and $N$ are finite everywhere\footnote{From the asymptotic behaviour of the $\elik$-function \citep{gradryz07}, the leading terms are:
\begin{equation}
\begin{cases}
M \sim - \zeta \ln \sqrt{(a-R)^2 + \zeta^2}\\\\
N \sim - (a-R) \ln \sqrt{(a-R)^2 + \zeta^2}
\end{cases}
\end{equation}
as $k \rightarrow 1$ (corresponding to $a \rightarrow R$ and $z  \rightarrow Z$).}, even onto the boundary ${\partial \cal S}$ where $k$ reaches unity, making the determination of the potential through Eq.(\ref{eq:curlt}) particularly simple and efficient.

We then use this nice and powerful result | reproduced in details in Appendix
\ref{app:mandn} | as the starting point of our investigation. By analogy,
we build the {\it incomplete form} of the axially symmetrical solution,
namely:
\begin{equation}
\label{eq:m0n0basis}
\begin{cases}
M_0 =\frac{1}{2}\zeta\sqrt{\frac{a}{R}} k F(\beta_0,k) + \frac{1}{2} f,\\\\
N_0 =\frac{1}{2} \sqrt{\frac{a}{R}} k \left\{ (a+R)  F(\beta_0,k) - \frac{2R}{k^2} \left[ F(\beta_0,k) - E(\beta_0,k) \right] \right\} + \frac{1}{2} g,
\end{cases}
\end{equation}
where $f$ and $g$ are unknown functions to be determined (but which must be zero under axial symmetry). The factor $\frac{1}{2}$ | not present in Eq.(\ref{eq:mandns0}) | comes from the fact that $M_1=-M_2$ under axial symmetry (see the end of Section \ref{sec:itg}). The same kind of strategy has been used in \citep{hure12}. This choice is also motivated by the fact that the partial derivatives of the incomplete elliptic integrals are very close to the partial derivatives of the complete ones. Actually, we have \citep{gradryz07}:
\begin{equation}
k {k'}^2 \partial_k F(\phi,k) = E(\phi,k)-{k'}^2 F(\phi,k)- \frac{k^2 \sin \phi \cos \phi}{\sqrt{1- k^2 \sin^2 \phi} },
\label{eq:partialk}
\end{equation}
and
\begin{equation}
k \partial_k E(\phi,k) = E(\phi,k)-F(\phi,k),
\label{eq:partiale}
\end{equation}
and so the differences between the complete version and the incomplete version come uniquely from the last term in Eq.(\ref{eq:partialk}) containing trigonometric functions. As we have
\begin{equation}
\zeta \partial_z k = k \left( 1 - \frac{k^2}{m^2} \right),
\label{eq:partialkz}
\end{equation}
where $(a+R)m=2\sqrt{aR}$, and
\begin{equation}
\partial_a k = \frac{1}{a+R} \zeta \partial_z k -\frac{k}{2a}\left(\frac{a-R}{a+R}\right),
\label{eq:partialka}
\end{equation}
we can easily take the partial derivatives of $M_0$ and $N_0$ defined above and set constraints upon the functions $f$ and $g$. After some algebra and rearrangement of terms, we find the general result:
\begin{equation}
\partial_a N_0 - \partial_z M_0 = \frac{1}{2}  \sqrt{\frac{a}{R}} k F(\beta_0,k) + \frac{1}{2}  \sqrt{\frac{R}{a}} \frac{k \sin \beta_0 \cos \beta_0}{\sqrt{1- k^2 \sin^2 \beta_0} } + \frac{1}{2}  \partial_a g - \frac{1}{2}  \partial_z f.
\label{eq:danmdzm}
\end{equation}
In order to satisfy Eq.(\ref{eq:mandns0}), we conclude that the sum of the last three terms in this relationship must vanish. As we have
\begin{equation}
\sqrt{\frac{R}{a}} \frac{k }{\sqrt{1- k^2 \sin^2 \beta_0} } =  \frac{2R}{\sqrt{(a+R)^2+\zeta^2 - 4aR  \sin^2 \beta_0}},
\end{equation}
it turns out that the following two basic pairs $(f,g)$ are appropriate:
\begin{equation}
\begin{cases}
f = - R \sin (2 \beta_0) \times \asinh \frac{\zeta}{\sqrt{(a+R)^2- 4aR \sin^2\beta_0}}\\\\
g = 0
\end{cases}
\end{equation}
or
\begin{equation}
\begin{cases}
f = 0\\\\
g = - R \sin (2 \beta_0) \times \asinh \frac{a+R \cos (2 \beta_0)}{\sqrt{\zeta^2 + R^2 \sin^2 (2\beta_0)}}
\end{cases}
\end{equation}
or {\it any linear combination}, for instance  of the form $\alpha f$ and  $(1-\alpha)g$ where $\alpha \in [0,1]$ (any function of $a$ can be added to $f$, and any function of $\zeta$ can be added to $g$). The optimal choice may depend on the actual problem. Note that $f$ and $g$ are regular (i.e. non-diverging) functions. Actually, $f=0$ as soon as i) $\zeta=0$ (i.e. in a horizontal plane containing the top/bottom of the cell), or ii) $\beta_0=\{0,\frac{\pi}{2}\}$ (i.e. along the line $\theta=\theta_0'$). For $a=R$ and $\beta_0 \rightarrow \frac{\pi}{2}$, we also have:
\begin{equation}
f \sim -2R \sin \beta_0 \cos \beta_0 \ln \frac{\zeta}{R | \cos \beta_0 |} \rightarrow 0.
\end{equation}

\begin{figure}
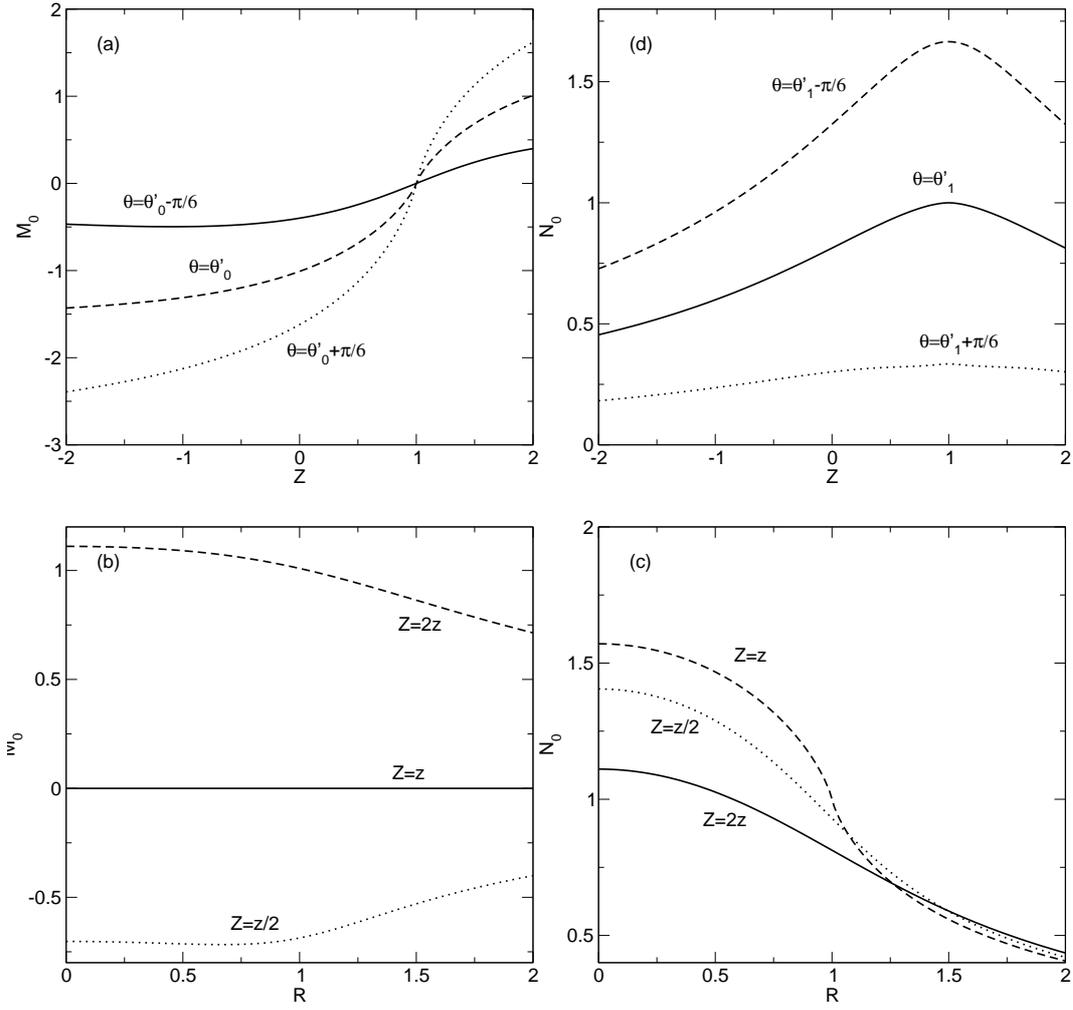

\centering
\includegraphics[width=7cm,bb=26 47 619 590,clip==]{mz.eps}\includegraphics[width=7cm,bb=26 47 619 590,clip==]{nz.eps}\\
\bigskip
\includegraphics[width=7cm,bb=26 47 619 590,clip==]{mr.eps}\includegraphics[width=7cm,bb=26 47 619 590,clip==]{nr.eps}
\caption{{\it Left panels}: variation of $M_0$ with $Z$ for $R=a$ and for three values of $\theta-\theta'$ ({\it a}), variation of $M_0$ with $R$ for $\theta=\theta'_0$ and three values of $Z$ ({\it b}); $\alpha=1$ in both cases (i.e. $f$ is included). {\it Right panels}: same but for $N_0$ and $\alpha=0$ (i.e. $g$ is included).}
\label{fig:m}
\end{figure}

The components of the generic problem are finally of the form:
\begin{equation}
\label{eq:m0n0}
\begin{cases}
M_0 =\frac{1}{2}\zeta\sqrt{\frac{a}{R}} k F(\beta_0,k) + \frac{1}{2} \alpha f,\\\\
N_0 = \frac{1}{2} \sqrt{\frac{a}{R}} k \left\{ (a+R)  F(\beta_0,k) - \frac{2R}{k^2} \left[ F(\beta_0,k) - E(\beta_0,k) \right] \right\}\\
\qquad \qquad + \frac{1}{2} (1-\alpha)g.
\end{cases}
\end{equation}
We have plotted in Fig. \ref{fig:m} (left panels) $M_0$ versus $Z$ for $R=a$ and $\theta-\theta_0'=\{-\frac{\pi}{6},0,+\frac{\pi}{6}\}$, and $M_0$ versus $R$ for $\theta=\theta_0'$ and $Z=\{\frac{z}{2},z,2z\}$. This computation includes the function $f$ with $\alpha=1$, and the parameters are $a=1$ and $z=1$. The function $N_0$ (including the function $g$ with $\alpha=0$) is plotted in the same figure (right panels). We see that  $M_0$ and $N_0$ are fully regular.

\begin{figure}
\includegraphics[width=6.5cm,bb=0 0 351 340,clip=]{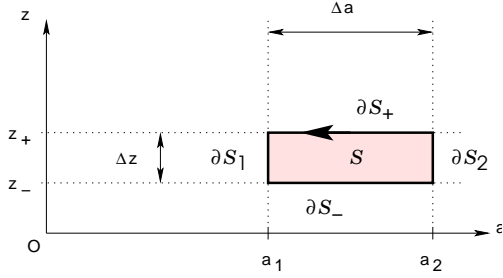}
\caption{Meridional cross-section $\cal S$ of the cylindrical cell.}
\label{fig:2Dcell}
\end{figure}

\section{The potential of a cylindrical cell}
\label{sec:cylcell}

From Eqs.(\ref{eq:curlt}) and (\ref{eq:m0n0}) together with the right expressions for $f$ and $g$, we can construct the potential due to a given cylindrical cell defined by $6$ parameters: $\beta_1$, $\beta_2$, $\ain$, $\aout$, $z_-$ and $z_+$ (see Fig. \ref{fig:possiblecells}a). Fortran 90 routines and a driver program which performs the numerical computation are available as a supplementary resource (see Appendix \ref{app:f90}). To remove any ambiguity, it is necessary to make explicit the dependency of all the functions with various parameters, i.e. $M_0 \equiv M_0(a,z,\beta_0)$, $N_0 \equiv N_0(a,z,\beta_0)$, $f \equiv f(a,z,\beta_0)$ and $g \equiv g(a,z,\beta_0)$. We still omit the position variables $R$ and $Z$ all these functions depend on, as well as the $\alpha$-parameter ($\alpha=1$ to include $f$ and exclude $g$, and $\alpha=0$ in the reverse case). The potential of the homogeneous cylindrical cell is:
\begin{equation}
\label{eq:psialongs}
\psi(\vec{r}) = -G \rho_0 \oint_{\partial \cal S}{Mda+Ndz},
\end{equation}
where
\begin{flalign}
\label{eq:M12}
M & \equiv M(a,z,\beta_1,\beta_2)\\ \nonumber
&  = M_1(a,z)-M_2(a,z),
\end{flalign}
and
\begin{flalign}
\label{eq:N12}
N & \equiv N(a,z,\beta_1,\beta_2)\\ \nonumber
&   = N_1(a,z)-N_2(a,z).
\end{flalign}

\begin{figure}
\centering
\includegraphics[width=8.5cm,bb=28 50 767 583,clip==]{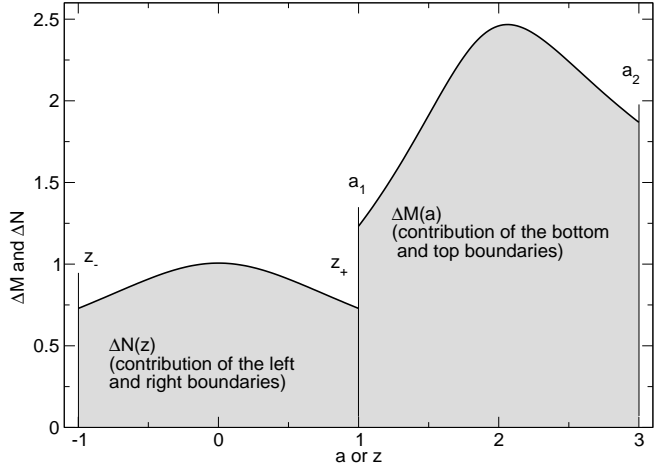}
\caption{$\Delta M(a)$ and $\Delta N(z)$ along the boundary ${\cal S}$ for P$(R,Z)=(2,0)$ and for the following cylindrical cell : $\theta_2'=-\theta_1'=+\frac{\pi}{12}$, $\ain=1$, $\aout=3$, $z_-=-1$ and $z_+=1$. The areas under the two curves (shown in grey) give, according to Eq.(\ref{eq:psi:alongs_rect}), the potential at P$(R,Z)$.}
\label{fig:mnovers}
\end{figure}

This formula is a line integral with regular integrands and is exact. As quoted
above, this holds even if the top and bottom edges are curved (i.e. if $a$
depends on $z$), as for instance considered in \cite{ansorg03}. In the present
case, $\partial {\cal S}$ is made of two rectangles, the one located at
$\theta_1'$, the other at $\theta_2'$, and each with corners at $(\ain,\zin)$,
$(\aout,\zin)$, $(\ain,\zout)$ and $(\ain,\zout)$, as shown in Fig.
\ref{fig:2Dcell}. To perform the integration along the whole boundary
(counter-clockwise), $\partial {\cal S}$ must be decomposed into four parts:
$\partial {\cal S}_-$ and $\partial {\cal S}_+$ for the bottom and top edges
($z=z_-$ and $z=z_+$), and $\partial {\cal S}_1$ and $\partial {\cal S}_2$ for
the left and right edges ($a=\ain$ and $a=\aout$). We then get the following
expression:
\begin{equation}
\psi(\vec{r})  =  -G \rho_0 \left[ \int_\ain^\aout{\Delta Mda} + \int_\zin^\zout{\Delta Ndz} \right].
\label{eq:psi:alongs_rect}
\end{equation}
where we have set:
\begin{flalign}
\label{eq:dm:alongs_rect}
\Delta M & \equiv M(a,\zin,\beta_1,\beta_2)-M(a,\zout,\beta_1,\beta_2)\\ \nonumber
& = M_1(\zin)+M_2(\zout)-M_2(\zin)- M_1(\zout),
\end{flalign}
which corresponds to $\partial {\cal S}_-$ and $\partial {\cal S}_+$, and
\begin{flalign}
\label{eq:dn:alongs_rect}
\Delta N & \equiv N(\aout,z,\beta_1,\beta_2)-N(\ain,z,\beta_1,\beta_2)\\ \nonumber
& = N_1(\aout)+ N_2(\ain)-N_2(\aout)-N_1(\ain),
\end{flalign}
which corresponds to $\partial {\cal S}_1$ and $\partial {\cal S}_2$. 
There is little chance that these integrations can be performed analytically to produce a closed-form, especially because of the presence of special functions. Figure \ref{fig:mnovers} displays $\Delta M$ versus $a$ and $\Delta N$ versus $z$ for the following cell parameters: $\ain=1$, $\aout=3$, $z_-=-1$, $z_+=1$, $\theta_1'=0$ and $\theta_2'=\frac{\pi}{6}$. The computation is performed at the centre of the cell (i.e. $R=2$, $Z=0$, $\theta=\frac{\pi}{12}$). The area under the curves $\Delta M(a)$ and $\Delta N(z)$ therefore yields the potential at this particular point.

\begin{figure}
\centering
\includegraphics[width=7.5cm,bb=0 0 432 361,clip=]{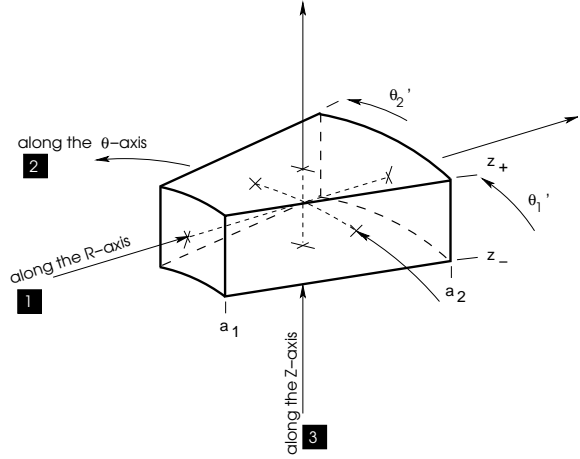}
\caption{The cylindrical cell and three different directions (see also Fig.\ref {fig:psi_3Dcell_123}).}
\label{fig:3Dcell}
\end{figure}

\begin{figure}
\centering
\includegraphics[width=8cm,bb=0 0 728 570,clip=]{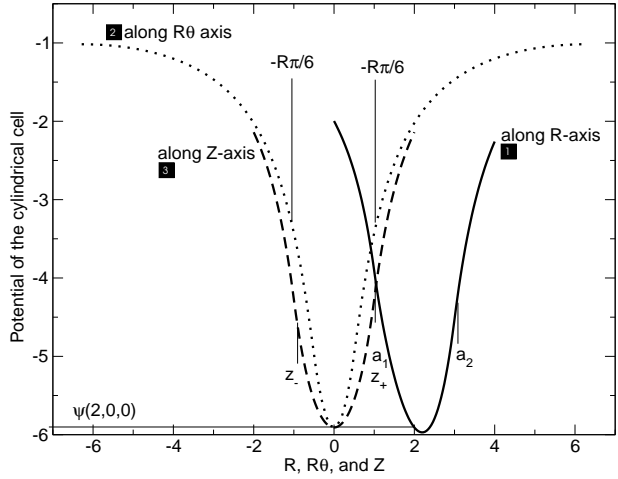}
\caption{The potential $\psi(R,\theta,Z)$ per unit mass of the cylindrical cell (same conditions as for Fig. \ref{fig:mnovers}) along three directions crossing-over the centre of the cell (see Fig. \ref{fig:3Dcell}) computed from Eq.(\ref{eq:psi:alongs_rect}): $R \in [0,4]$ for $\theta=0$ and $Z=0$, $R\theta \in [0,2\pi R]$ for $R=2$ and $Z=0$, and $Z \in [-2,2]$ for  $R=2$ and $\theta=0$. These directions are shown in Fig. \ref{fig:3Dcell}. The potential at the geometrical centre is $\psi(2,0,0) \approx -5.90798$. }
\label{fig:psi_3Dcell_123}
\end{figure}


\begin{figure}
\centering
\hspace*{-1cm}\includegraphics[width=6.5cm,bb=151 80 554 500,clip=,angle=-90]{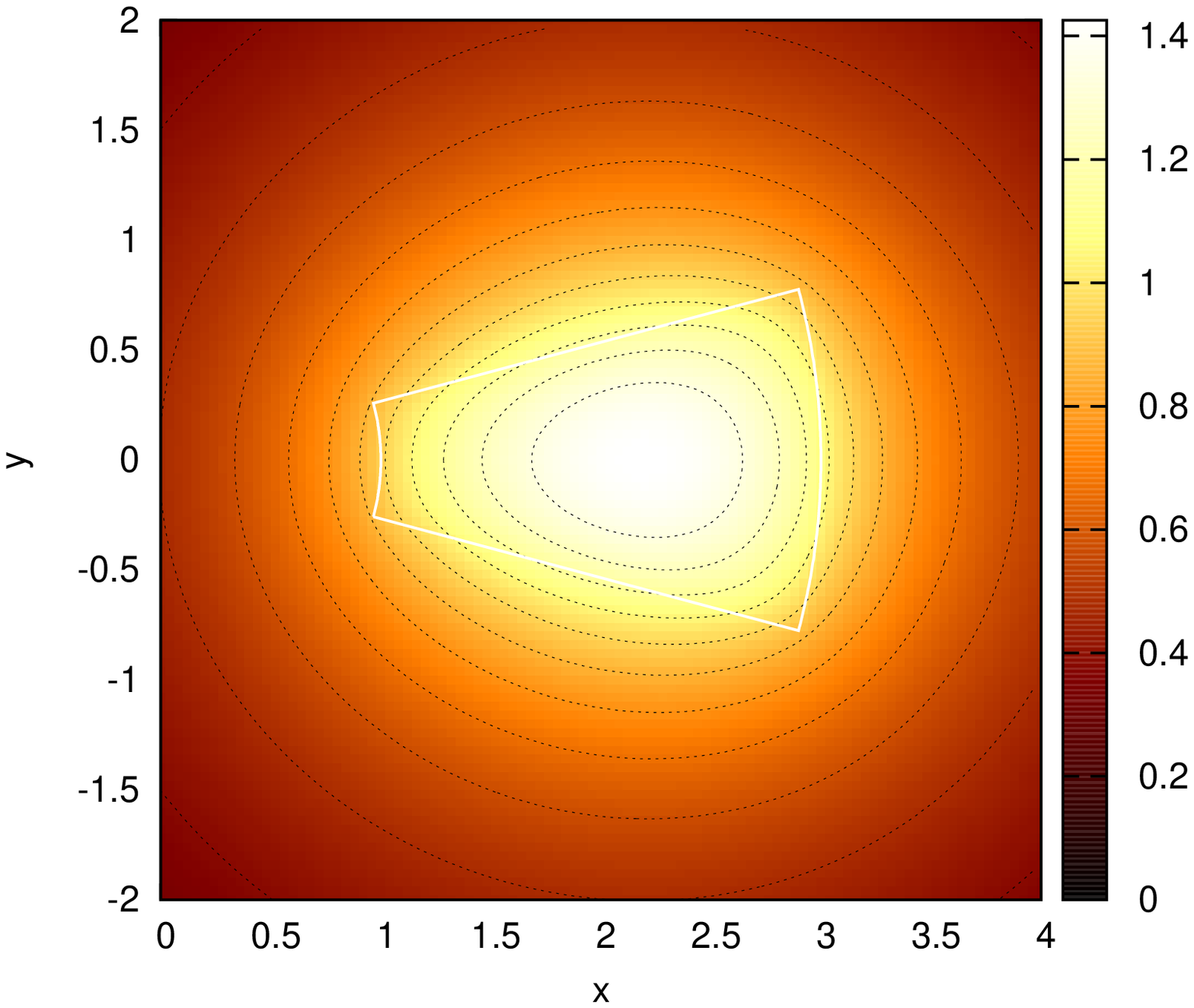}\includegraphics[width=6.5cm,bb=151 112 554 554,clip=,angle=-90]{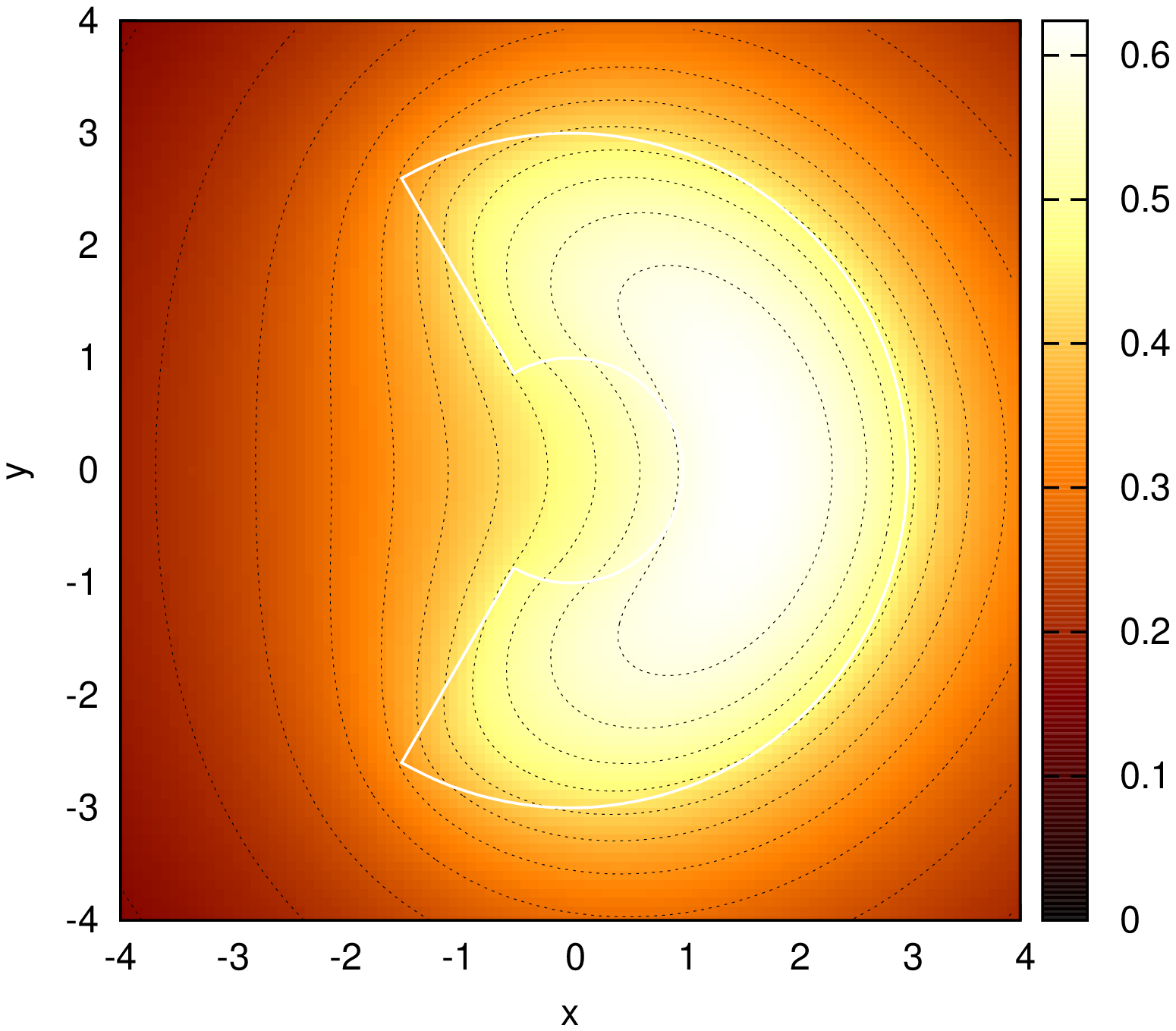}\\
\hspace*{-1cm}\includegraphics[width=6.5cm,bb=151 80 554 500,clip=,angle=-90]{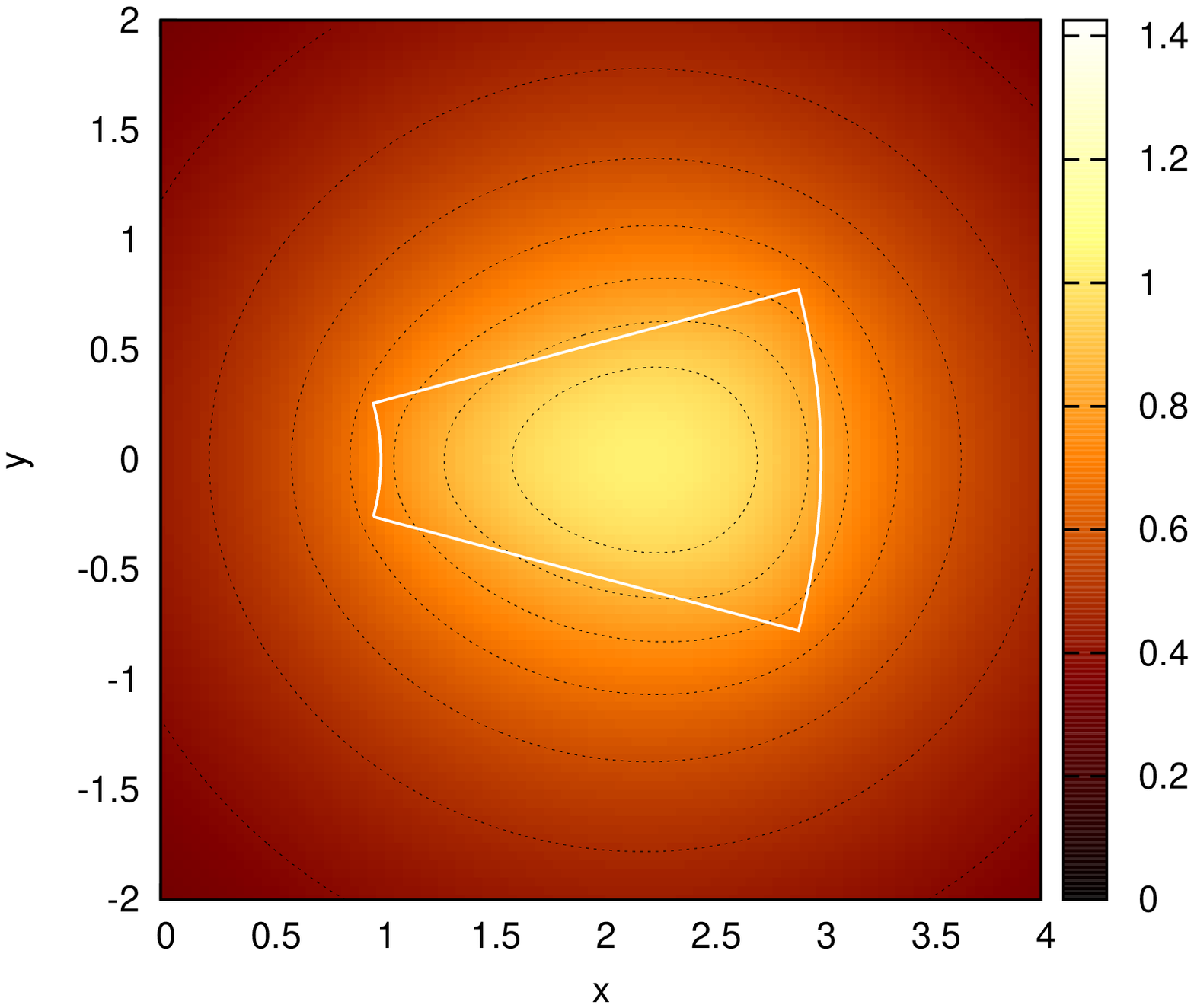}\includegraphics[width=6.5cm,bb=151 112 554 554,clip=,angle=-90]{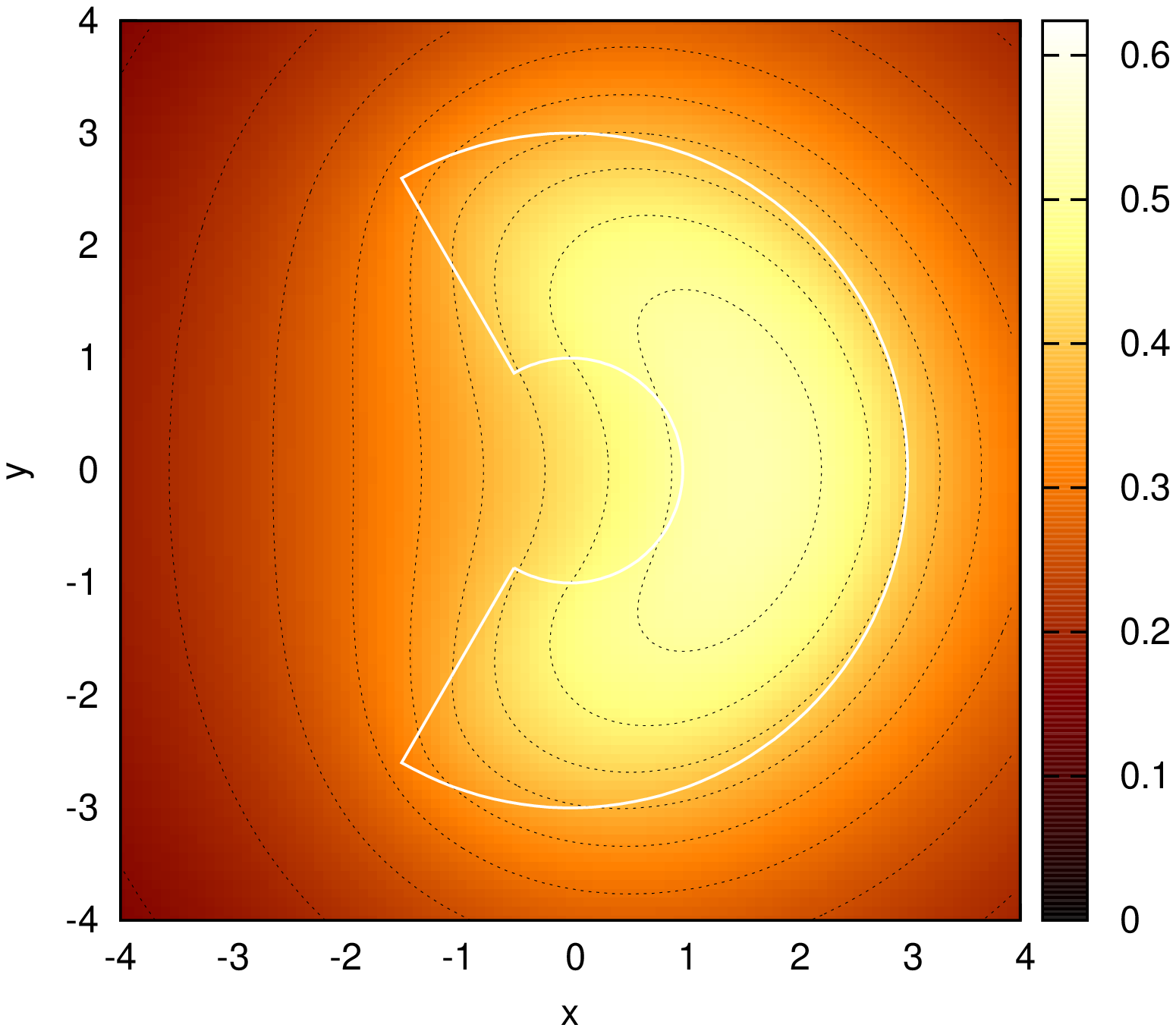}\\
\hspace*{-1cm}\includegraphics[width=6.5cm,bb=151 80 554 500,clip=,angle=-90]{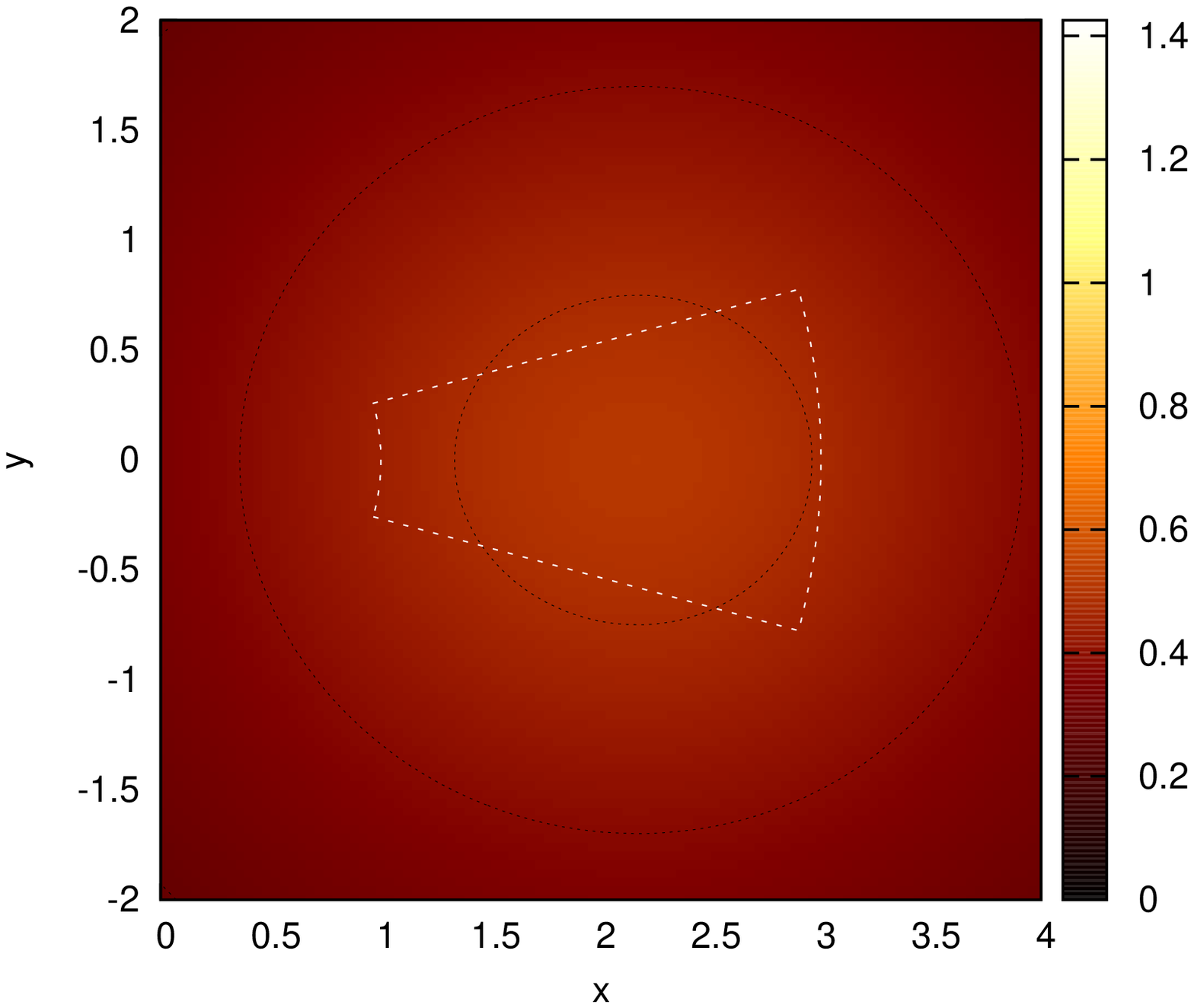}\includegraphics[width=6.5cm,bb=151 112 554 554,clip=,angle=-90]{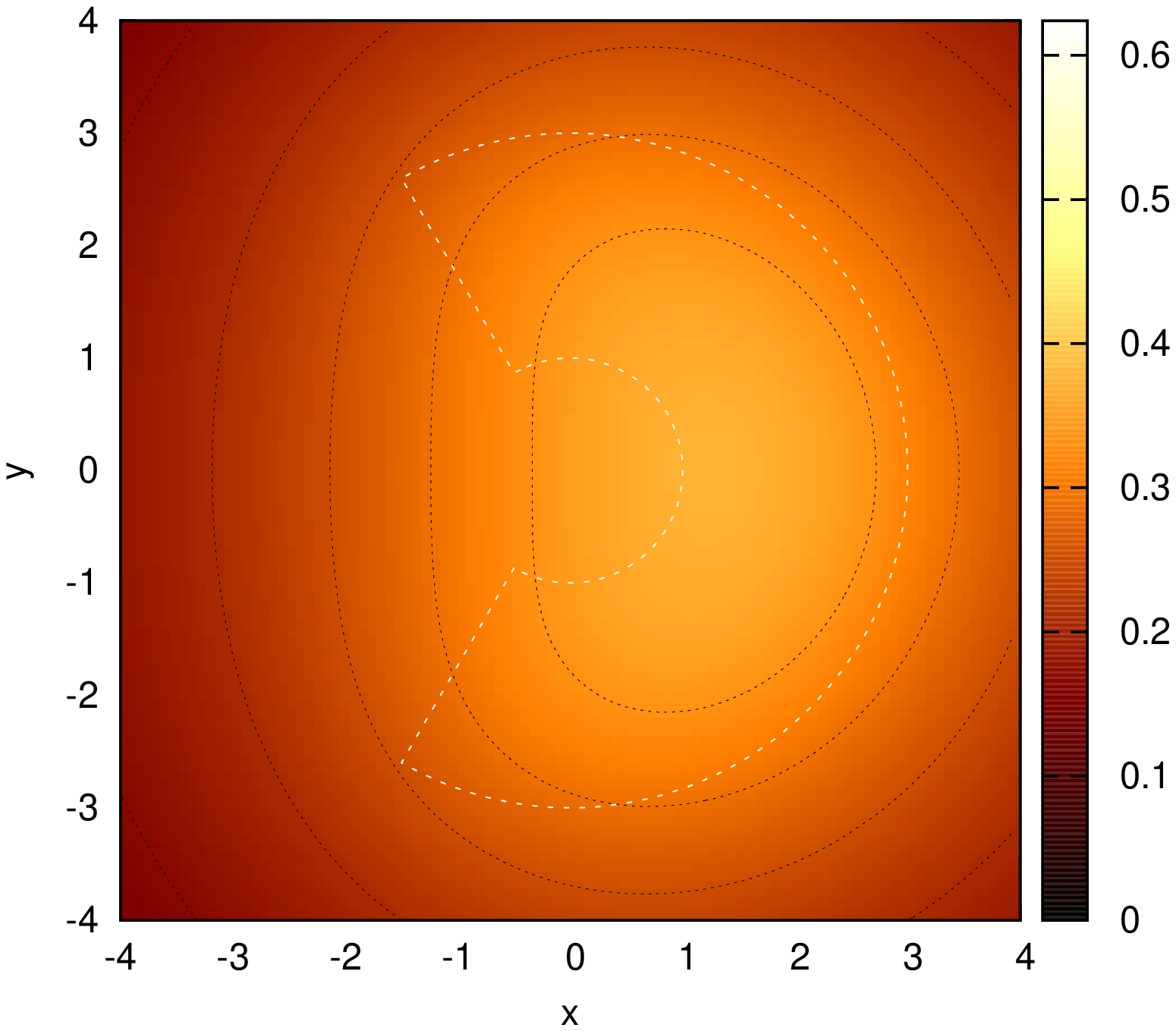}
\caption{{\it Left panels}: potential per unit mass $\psi(R,\theta,Z)/{\cal M}$ of the cylindrical cell (i.e. same conditions as for Fig. \ref{fig:mnovers} in three horizontal planes : $Z=0$ ({\it top panel}), $Z=z_+$ ({\it middle panel}) and $Z=2z_+$ ({\it bottom panel}). The colour code is the same. The section of the cell is shown in white. The total mass is ${\cal M} \approx 4.19$. {\it Right panels}: same but with $\theta_2'=-\theta_1'=+\frac{2\pi}{3}$. The total mass is  ${\cal M} \approx 33.5$.}
\label{fig:psi_3Dcell_full1}
\end{figure}

The numerical integration of $\Delta M(a)$ and $\Delta N(z)$ is easy to perform and contains no pitfall. Here, we use the second-order trapezoidal rule with $n$ nodes per direction. Figure \ref{fig:3Dcell} displays $3$ preferential directions crossing-over the cylindrical cell, namely:
\begin{itemize}
\item direction $1$, corresponding to the radial direction ($R$ varies while $\theta$ and $Z$ are held fixed),
\item direction $2$, corresponding to the azimuthal direction ($\theta$ varies while $R$ and $Z$ are held fixed),
\item direction $3$, corresponding to the vertical direction ($Z$ varies while $R$ and $\theta$ are held fixed).
\end{itemize}

The potential $\psi(R,\theta,Z)$ has been determined along these $3$-directions from Eq.(\ref{eq:psi:alongs_rect}) using the basic, trapezoidal rule (the aim is not to optimise the accuracy which can obviously be easily increased by using other schemes). It is shown in Fig. \ref{fig:psi_3Dcell_123}. We see that the potential has finite gradients when crossing-over the boundary of the cell, which is typical of matter continuously distributed in three dimensions. The curvature effect is clearly visible from the asymmetrical variation of $\psi$ with $R$ (direction $1$). Left panels in Fig. \ref{fig:psi_3Dcell_full1} displays the potential of the cell in the $Z=0$, $Z=z_+$ and $Z=2z_+$ planes. Right panels are for an opening angle $\Delta \theta' = +\frac{4\pi}{3}$. Values are given by unit of mass
\begin{equation}
{\cal M} = \rho_0 \Delta z \Delta a a_0 \Delta \theta',
\end{equation}
where $\Delta z=z_+-z_-$ is the vertical extension, $\Delta a = \aout - \ain$ is the radial extension, $\Delta \theta' =\theta_2'-\theta_1'$ is the opening angle and $a_0 = \frac{1}{2}(\aout + \ain)$ is the mean radius.\\

\begin{figure}
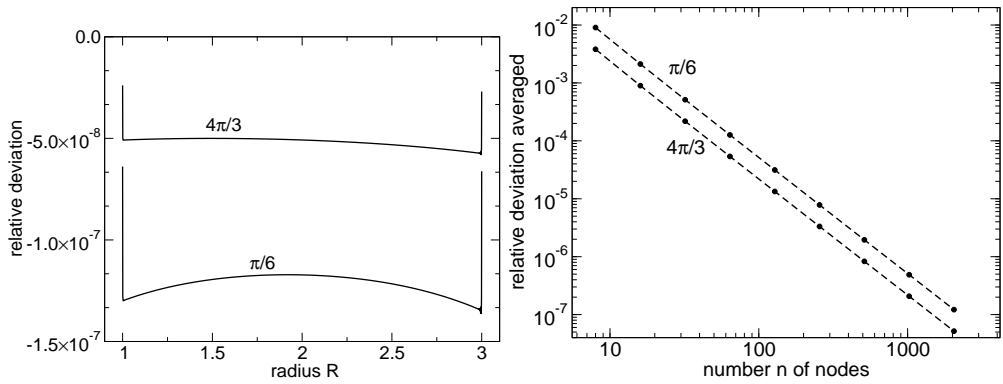

\centering
\includegraphics[width=6.5cm,bb=17 40 706 531,clip==]{cmp2approaches.eps}\;\includegraphics[width=6.5cm,bb=70 39 706 522,clip==]{cmp2vsn.eps}
\caption{{\it Left}: Relative deviation versus the radius between potential values computed from the contour integral (this work) and computed by vertically integrating the potential of the polar cell, for the two cells considered in Fig. \ref{fig:psi_3Dcell_full1}. The computation is here limited to part of the boundary with $\theta=\theta'_1$ and $Z=z_-$. {\it Right}: averaged deviation versus the number $n$ of nodes used to perform the numerical quatratures.}
\label{fig:cmp2approaches.eps}
\end{figure}

It happens that there is a closed form | not reported here | for the integration over $z$ in Eq.(\ref{eq:psi3}) in the homogeneous case, which corresponds to the potential of a polar cell \citep{hure12}. Thus, the potential of a cylindrical cell can also be obtained numerically by integrating that formula along the $z$-direction and it is expected to perfectly match the contour integral given by Eq.(\ref{eq:psialongs}). We can therefore check the correctness of the contour integral relative to direct integration. This is shown in Fig. \ref{fig:cmp2approaches.eps} where we have plotted the relative deviation versus the radius $R$ between the two potential values computed along a quarter of the boundary (i.e., for  $\theta=\theta'_1$ and $Z=z_-$). We have considered here two different opening angles, $\Delta \theta' \in \{\frac{\pi}{6},\frac{4\pi}{3}\}$, as for Figure \ref{fig:psi_3Dcell_full1}. The averaged deviation computed along the boundary versus the number of node $n$ used for the quadratures is shown on the second panel. We see that the two approaches give the same results, the accuracy being limited by the quadrature scheme, which is second-order. In fact, the number of evaluations of special functions is almost the same (although there is no need here for the elliptic integral of the third kind here), and the equation for the boundary $z(a)$ is required in both cases. 

Finally, acceleration is easily deduced from $\psi$, and we see from Eq.(\ref{eq:psialongs}) that the acceleration vector is also defined by a line integral, namely
\begin{flalign}
\nonumber
- \vec{\nabla} \psi & =  G  \rho_0  \vec{\nabla}  \oint_{\partial \cal S}{Mda+Ndz},\\
& =  G  \rho_0 \oint_{\partial \cal S}{ \vec{\nabla}Mda+\vec{\nabla}Ndz}.
\label{eq:acc:alongs_rect}
\end{flalign}

To get the three cylindrical components, we therefore need to calculate the partial derivatives of $M$ and $N$ with respect to $R$, $\theta$ and $Z$. The final formulae are accessible from the gradients of $M_0$ and $N_0$ which are reported in the appendix \ref{sec:accs}. We however note a difficulty here: the acceleration at the edges of the polar cell is logarithmically singular and the formulae given in the appendix have the same drawback. A closed form would really be needed. In the meanwhile, it is always possible to determine the acceleration of the $3$D-cell | a finite vector | from the potential through finite differences, which is technically easy.

\section{Conclusion}

In this article, we have shown that the gravitational potential of elementary
cells which compose certain polar-type grids can be determined from a one
dimensional integral. This result generalises an integral expression already
known under axial symmetry \citep{hw53,ansorg03} and enables to treat fully
inhomogeneous systems. The new formula is a contour integral which depends on
the shape of the grid cells. It means that we can compute it once for all. This
is quite advantageous when working in a fixed cylindrical grid. As its kernel
is regular, it can be easily computed by standard means with great accuracy.
This is especially interesting if one wishes to generate reference solutions
for a given cell's geometry. In particular, we have applied the integral formula to the case of a cylindrical cell. The approach can be generalised to other shapes, as soon as they are well bounded in the
azimuthal direction. Such objects include the spherical cell, as well as
objects with variable vertical thickness as a function of the radius (sectors
of tori or flared disks, etc.).

\begin{acknowledgements}

We greatly thank D. Pfenniger for his suggestions about a preliminary version of the project. We thank the anonymous referees for their valuable comments and suggestions to improve the presentation.

\end{acknowledgements}


\appendix

\section{Values of $F(\beta,k)$ and $E(\beta,k)$ for any amplitude $\beta$}
\label{app:f}

The computation of $F(\beta,k)$ must be performed with caution as soon as the amplitude $\beta$ stands outside the range $[0,\frac{\pi}{2}]$. It is in particular necessary to use the following rules:
\begin{equation}
\begin{cases}
F(\beta,k)=-F(-\beta,k), \quad \mathrm{if \;} \beta < 0,\\\\
F(\beta,k)=n \elik(k)+F\left(|\beta|-n\frac{\pi}{2},k\right), \quad \mathrm{if \;} |\beta|-n\frac{\pi}{2} \in [0,\frac{\pi}{2}] \mathrm{\; and \; even \;} n,\\\\
F(\beta,k)=n \elik(k)-F\left(n\frac{\pi}{2}-|\beta|,k\right),\quad \mathrm{if \;} n\frac{\pi}{2}-|\beta|\in [0,\frac{\pi}{2}] \mathrm{\; and \; odd \;} n,
\end{cases}
\end{equation}
and
\begin{equation}
\begin{cases}
E(\beta,k)=-E(-\beta,k), \quad \mathrm{if \;} \beta < 0,\\\\
E(\beta,k)=n \elie(k)+E\left(|\beta|-n\frac{\pi}{2},k\right), \quad \mathrm{if \;} |\beta|-n\frac{\pi}{2} \in [0,\frac{\pi}{2}] \mathrm{\; and \; even \;} n,\\\\
E(\beta,k)=n \elie(k)-E\left(n\frac{\pi}{2}-|\beta|,k\right), \quad \mathrm{if \;} n\frac{\pi}{2}-|\beta|\in [0,\frac{\pi}{2}] \mathrm{\; and \; odd \;} n.
\end{cases}
\end{equation}

\section{Derivation of $M$ and $N$ in the axially symmetrical case}
\label{app:mandn}

In order to determine $M$ and $N$ in the following equation
\begin{equation}
\partial_a N - \partial_z M = 2 \sqrt{\frac{a}{R}} k \elik(k),
\end{equation}
we set without loss of generality:
\begin{equation}
\begin{cases}
M = \zeta k \elie(k) f(a) +  \zeta k \elik(k) g(a),\\\\
N = a k \elie(k) h(a,\zeta) +  a k \elik(k) l(a,\zeta),
\end{cases}
\end{equation}
where $f$, $g$, $h$ and $l$ are four functions to be determined. From Eqs.(\ref{eq:partialkz}) and (\ref{eq:partialka}), and given \citep{gradryz07}:
\begin{equation}
\partial_k k\elie(k) = 2 \elie(k) - \elik(k),
\end{equation}
and
\begin{equation}
\partial_k k\elik(k) = \frac{\elie(k)}{{k'}^2},
\end{equation}
we get:
\begin{flalign}
-\partial_z M & = k\elie(k) f(a) + k\elik(k) g(a) +  \zeta \partial_\zeta k \left[ \left( 2\elie(k) - \elik(k) \right) f +  g \frac{\elie(k)}{{k'}^2}  \right] \\
\nonumber
& = k\elie(k) \left[ \left( 2\frac{k^2}{m^2}-1\right) f - \left( 1 - \frac{k^2}{m^2} \right) \frac{g}{{k'}^2} \right]  +  k\elik(k)   \left[ \left( 1 - \frac{k^2}{m^2} \right) f + g \right]
\nonumber
\end{flalign}
and
\begin{flalign}
\partial_a N & = k\elie(k) (ah' + h) +  k\elik(k) (al'+l) +  a \partial_a k \left[ \left( 2\elie(k) - \elik(k) \right) h +  l \frac{\elie(k)}{{k'}^2}  \right] \\
\nonumber
& = k\elie(k) \left[ ah'+h + \left(2h + \frac{l}{{k'}^2} \right) \left( \frac{1}{2} - \frac{a}{a+R}\frac{k^2}{m^2} \right) \right]\\\nonumber
& \qquad +  k\elik(k) \left[ al'+l  - h \left( \frac{1}{2} - \frac{a}{a+R} \frac{k^2}{m^2} \right) \right].
\end{flalign}

Forming $\partial_a N-\partial_z M$, and gathering terms, we get 
\begin{flalign}
 2h + ah' - f - \frac{g}{{k'}^2} + \frac{l}{{2k'}^2} + \frac{k^2}{m^2} \left( 2f + \frac{g}{{k'}^2} -\frac{2a}{a+R}h - \frac{l}{{k'}^2}\frac{a}{a+R} \right)
\end{flalign}
for the term multiplying $k\elie(k)$, and
\begin{equation}
f+g+l+al'-\frac{h}{2}+\frac{k^2}{m^2} \left( h \frac{a}{a+R}-f \right)
\end{equation}
for the term multiplying $k\elik(k)$. The solution by \citep{ansorg03} which corresponds to
\begin{equation}
\begin{cases}
M = \zeta \sqrt{\frac{a}{R}} k \elik(k)\\\\
N= \sqrt{\frac{a}{R}} k \left\{ (a+R)  \elik(k) - \frac{2R}{k^2} \left[  \elik(k) -   \elie(k) \right] \right\},
\end{cases}
\end{equation}
 is obtained for the following settings:
\begin{equation}
\begin{cases}
f=0,\\\\
g=\sqrt{\frac{a}{R}},\\\\
ah=\sqrt{\frac{a}{R}} \frac{2R}{k^2},\\\\
al=\sqrt{\frac{a}{R}} \left(a+R-\frac{2R}{k^2}\right) = \frac{1}{2}\sqrt{\frac{a^3}{R}}- \frac{R^2+\zeta^2}{2\sqrt{aR}},
\end{cases}
\end{equation}
which eliminates the term $k \elie(k)$ and produces the factor $2 \sqrt{\frac{a}{R}}$ for the term $k\elik(k)$.

\section{A basic Fortran 90 program}
\label{app:f90}

Fortran 90 routines and a driver program which computes
the $M$ and $N$ functions for a cylindrical cell and the associated potential at
one space point $(R,\theta,Z)$ are available from the online version of the paper. The quadrature is performed from a Newton-Cotes, second-order quadrature scheme. External calls to functions {\tt
IEF(BETA,K)} and {\tt IEE(BETA,K)} refer to the values of the incomplete
elliptic integral $F(\beta,k)$ and $E(\beta,k)$ respectively which can be
obtained from any mathematical library. For single (double) precision
computations, change {\tt \_AP} into {\tt 4} (resp  {\tt 8}), and {\tt EPSMACH}
is the corresponding precision (about $2 \times 10^{-16}$ in double precision).
These routines are not optimized. Running the code with the default parameters
generates the following output:
{\small
\begin{verbatim}
 A0,A1,A2,DELTAA   1.0000000      0.75000000       1.2500000      0.50000000    
 Z0,Z1,Z2,DELTAZ   0.0000000     -0.25000000      0.25000000      0.50000000    
 THETA0,THETA1,THETA2,DELTATHETA   0.0000000     -0.25000000      0.25000000      0.50000000    
 MASS  0.12500000    
 R,THETA,Z,POT   1.0000000       0.0000000       0.0000000     -0.59440136    
\end{verbatim}
}

\section{Accelerations}
\label{sec:accs}

 Since $M$ and $N$ are build from the two elementary kernels $M_0$ and $N_0$ from Eq.(\ref{eq:m0n0}), we will only expand  $\partial_R M_0$, $\partial_R N_0$, $\partial_\theta M_0$, etc. After calculus, we get
\begin{equation}
\begin{cases}
\frac{\partial  M_0 }{\partial R} = - \frac{1}{4}\sqrt{\frac{a}{R^3}} \zeta k^3 \left\{ \left[E(\beta_0,k) - \frac{k^2 \sin^2 \beta_0}{2 \sqrt{1-k^2\sin^2 \beta_0}} \right] \right.\\
\left.\qquad\qquad \times \left[1-\frac{(a+R)k^2}{2a} \right]\frac{1}{{k'}^2}+F(\beta_0,k) \right\} + \frac{1}{2} \alpha \frac{\partial  f }{\partial R}\\
\frac{\partial  N_0 }{\partial R} = - \frac{1}{4}\sqrt{\frac{a}{R^3}} k \frac{a^2+R^2}{a}F(\beta_0,k) + \frac{1}{2}\frac{kE(\beta_0,k)}{{k'}^2} \left\{\frac{a-R}{a+R}\sqrt{\frac{a}{R}}\left(\frac{a+R}{2R}-\frac{k^2}{m^2}\right)-\frac{{k'}^2}{m}\right\}\\
\qquad\qquad + \frac{1}{4}k \sqrt{\frac{a}{R}} \left(1-\frac{k^2}{m^2}+\frac{a-R}{2R} \right) \left({k'}^2-\frac{a-R}{a+R}\right) \frac{\sin(2\beta_0)}{{k'}^2\sqrt{1-k^2\sin^2 \beta_0}}\\
\qquad\qquad + \frac{1}{2}(1-\alpha) \frac{\partial  g }{\partial R}\\
\frac{\partial  M_0 }{\partial \theta} = \frac{1}{4}\sqrt{\frac{a}{R}}  \zeta \frac{k}{\sqrt{1-k^2\sin^2 \beta_0}} + \frac{1}{2} \alpha \frac{\partial  f }{\partial \theta}\\
\frac{\partial  N_0 }{\partial \theta} = \frac{1}{4}\sqrt{\frac{a}{R}} [a+R \cos (2 \beta_0)] \frac{k}{\sqrt{1-k^2\sin^2 \beta_0}} +  \frac{1}{2}(1-\alpha) \frac{\partial  g }{\partial \theta}\\
\frac{\partial  M_0 }{\partial Z} =\frac{1}{2}\sqrt{\frac{a}{R}}k \left[F(\beta_0,k)-\left(1-\frac{k^2}{m^2}\right)\frac{E(\beta_0,k)}{{k'}^2}\right]\\
\qquad\qquad + \frac{1}{4}\sqrt{\frac{R}{a}}\left(1-\frac{k^2}{m^2}\right)\frac{k^3}{{k'}^2\sqrt{1-k^2\sin^2 \beta_0}} + \frac{1}{2}\alpha \frac{\partial  f }{\partial Z}\\
\frac{\partial  N_0 }{\partial Z} = - \frac{1}{2}\sqrt{\frac{a}{R}}\left\{\left(a+R-\frac{2R}{k^2}\right)\left[E(\beta_0,k)-\frac{k^2 \sin(\beta_0)}{2\sqrt{1-k^2\sin^2 \beta_0}}\right]\right.\\
\qquad\qquad \left.+\frac{2R{k'}^2}{k^2}F(\beta_0,k)\right\}\frac{k}{\zeta{k'}^2}\left(1-\frac{k^2}{m^2}\right) + \frac{1}{2}(1-\alpha)\frac{\partial  g }{\partial Z}
\end{cases}
\end{equation}
where
\begin{equation}
\begin{cases}
\frac{\partial  f }{\partial R} = \frac{f}{R}+ \frac{1}{2} \sin (2 \beta_0) \sqrt{\frac{R}{a}} \frac{1}{1-m^2 \sin^2 \beta_0}\frac{k}{\sqrt{1-k^2\sin^2 \beta_0}} \frac{\zeta [R+a \cos (2 \beta_0)]}{(a+R)^2}\\\\
\frac{\partial  g }{\partial R} = \frac{g}{R}-  \frac{1}{2} \sin (2 \beta_0)  \sqrt{\frac{R}{a}} \frac{k}{\sqrt{1-k^2\sin^2 \beta_0}}  \frac{[\zeta^2 \cos(2\beta_0)-aR \sin^2(2\beta_0)]}{\zeta^2+R^2 \sin^2 (2\beta_0)} \\
\frac{\partial  f }{\partial \theta} = -R \cos (2 \beta_0) \asinh \frac{\zeta}{(a+R)\sqrt{1-m^2 \sin^2 \beta_0}} \\
\qquad\qquad - \frac{1}{2} R \sin (2 \beta_0) \frac{\zeta \sqrt{aR}}{(a+R)^2}\frac{ \sin(2\beta_0)}{1-m^2 \sin^2 \beta_0}\frac{k }{\sqrt{1-k^2\sin^2 \beta_0}}\\
\frac{\partial  g }{\partial \theta} = -R \cos (2 \beta_0) \asinh \frac{a+R \cos(2\beta_0)}{\sqrt{\zeta^2+R^2 \sin^2 (2\beta_0)}} \\
\qquad\qquad +  \frac{1}{2} R \sin (2 \beta_0)  \sqrt{\frac{R}{a}}\frac{[\zeta^2+R^2 +(a+R)R \cos(2\beta_0)] \sin(2\beta_0)}{\zeta^2+R^2 \sin^2 (2\beta_0)}\frac{k}{\sqrt{1-k^2\sin^2 \beta_0}} \\
\frac{\partial  f }{\partial Z} = -\frac{1}{2} \sin (2 \beta_0) \sqrt{\frac{R}{a}} \frac{k}{\sqrt{1-k^2\sin^2 \beta_0}}\\
\frac{\partial  g }{\partial Z} = + \frac{1}{2} \sin (2 \beta_0) \sqrt{\frac{R}{a}} \frac{k}{\sqrt{1-k^2\sin^2 \beta_0}} \frac{\zeta[a+R\cos(2\beta_0)]}{\zeta^2+R^2 \sin^2 (2\beta_0)}
\end{cases}
\end{equation}

\end{document}